\documentstyle[prb,aps,epsfig]{revtex}

\begin{document}
\draft
\title{Imaging of microwave permittivity, tunability, and damage recovery in (Ba,Sr)TiO$_{3}$ thin films}
\author{D. E. Steinhauer,$^{a)}$\footnotetext{$^{a)}$
Electronic mail: steinhau@squid.umd.edu. Color versions of the figures in
this paper can be found at http://www.csr.umd.edu/research/hifreq/micr\_microscopy.html.} C. P. Vlahacos, F. C. Wellstood, and Steven M. Anlage}
\address{Center for Superconductivity Research, Department of Physics, University of Maryland, College Park, MD 20742-4111 }
\author{C. Canedy and R. Ramesh}
\address{Department of Materials and Nuclear Engineering, University of Maryland, College Park, MD 20742-2115}
\author{A. Stanishevsky and J. Melngailis} 
\address{Department of Electrical Engineering, University of Maryland, College Park, MD 20742-3285}
\maketitle

\begin{abstract}
We describe the use of a near-field scanning microwave microscope to quantitatively image the dielectric permittivity and tunability of thin-film dielectric samples on a length scale of 1 $\mu$m.  We demonstrate this technique with permittivity images and local hysteresis loops of a 370 nm thick Ba$_{0.6}$Sr$_{0.4}$TiO$_{3}$ thin film at 7.2 GHz.
We also observe the role of annealing in the recovery of dielectric tunability in a damaged region of the thin film.
We can measure changes in relative permittivity $\epsilon_{r}$ as small as 2 at $\epsilon_{r} = 500$, and changes in dielectric tunability $d\epsilon_{r}/dV$ as small as 0.03 $V^{-1}$.
\end{abstract}

Many techniques exist for quantitatively measuring the dielectric permittivity of thin film samples.  For example, thin film capacitors allow measurement of the in-plane\cite{Hoerman,WChang} and normal\cite{QXJia,Miranda} components of the permittivity tensor.  Dielectric resonators\cite{Findikoglu} and Corbino measurements\cite{Booth} have also been used.  However, thin film capacitors require the sample to be altered, possibly affecting the dielectric permittivity; most other methods either require bulk samples or have poor spatial resolution.  Recently, near-field techniques, based on coaxial cavity\cite{Cho,Gao,Chang} microwave probes which use the reflected microwave signal from a sample, have been  used to determine the dielectric constant $\epsilon_{r}$ and tunability (i.e., how $\epsilon_{r}$ depends on voltage) of thin films.  However, no quantitative tunability measurements at high spatial resolution have been demonstrated.  In this letter, we present a nondestructive, noninvasive near-field scanning microwave microscope\cite{Steinhauer1,Steinhauer2,AnlageASC,VlahacosDielectric} which can quantitatively image the local permittivity and tunability of dielectric thin films with a spatial resolution of 1 $\mu$m.

\begin{figure}
\begin{center}
\leavevmode
\epsfxsize=10cm
\epsfbox{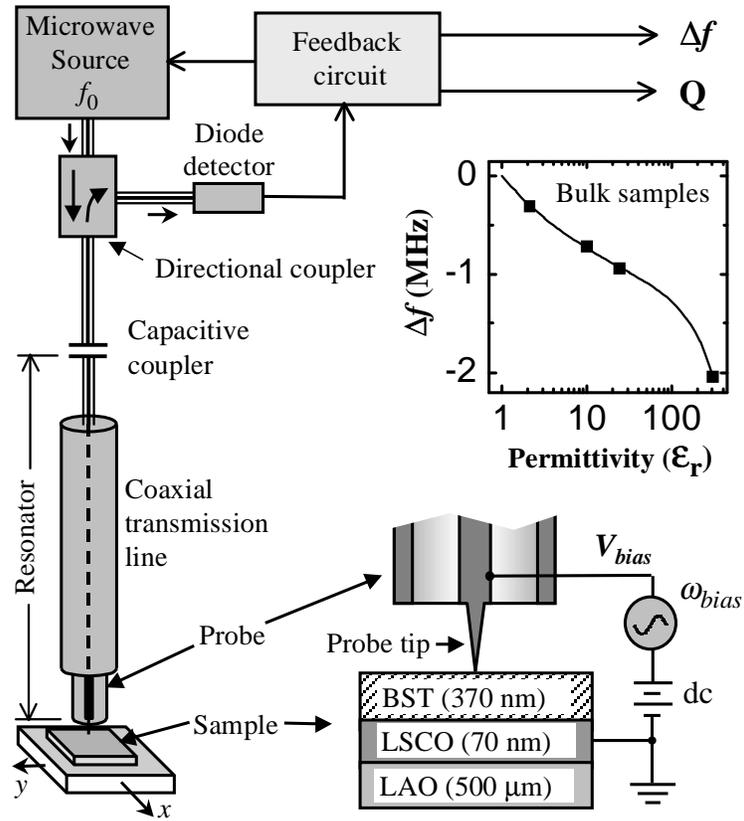}
\end{center}
\caption{Schematic of the near-field scanning microwave microscope.  The open-ended coaxial probe has a sharp tip (see inset) which is held in gentle contact with the sample.  The graph shows the frequency shift ($\Delta f$) as a function of dielectric permittivity ($\epsilon_{r}$) for a series of 500 $\mu$m thick bulk samples at 7.2 GHz.  The data points indicate experimental results, while the line shows model results.}
\label{schematic}
\end{figure}

Our near-field scanning microwave microscopes consist of an open-ended coaxial probe with a sharp, protruding center conductor (Fig.\ \ref{schematic}).  The probe is connected to a coaxial transmission line resonator, which is coupled to a microwave source through a capacitive coupler.  The probe tip, which has a radius $\sim 1\ \mu$m (see inset to Fig.\ \ref{schematic}), is held fixed, while the sample is supported by a spring-loaded cantilever applying a controlled normal force of about 50 $\mu$N between the probe tip and the sample. Due to the concentration of the microwave fields at the tip, the boundary condition of the resonator, and hence, the resonant frequency $f_{0}$ and quality factor Q, are perturbed depending on the dielectric properties of the region of the sample immediately beneath the probe tip.  We monitor this perturbation with a feedback circuit\cite{Steinhauer1,Steinhauer2} which receives the reflected microwave signal from the resonator, keeps the microwave source locked onto the chosen resonance, and outputs the frequency shift ($\Delta f$) and Q of the resonator. We have shown that the spatial resolution of the microscope in this mode of operation is about 1 $\mu$m\cite{AnlageASC}. In addition, a local dc electric field can be applied to the sample by means of a bias tee in the resonator.  

To observe the microscope's response to sample dielectric permittivity $\epsilon_{r}$, we monitored the frequency shift signal while scanning samples with known $\epsilon_{r}$.  With well-characterized 500 $\mu$m thick bulk dielectrics, we observed that the microscope frequency shift monotonically increases in the negative direction with increasing sample permittivity (see graph inset, Fig.\ \ref{schematic}).\cite{VlahacosDielectric}  We observed similar behavior with thin film samples.  

For comparison, we did a finite element calculation of the rf electric field near the probe tip.  Because the probe tip radius is much less than the wavelength $(\lambda \sim 4$ cm at 7 GHz), a static calculation of the electric field is sufficient.  Cylindrical symmetry further simplifies the problem to two dimensions.  We represent the probe tip as a cone with a blunt end, held at a potential of $\phi = 1$ V.  Using relaxation methods\cite{numericalrecipes} we solved Poisson's equation for the potential, $\nabla ^{2}\phi = 0$, on a rectangular grid representing the region around the probe tip.  The two variables we used to represent the properties of the probe were the aspect ratio of the probe tip ($\alpha \equiv dz/dr$) and the radius $r_{0}$ of the blunt end.

Since the sample represents a small perturbation to the resonator, we can use perturbation theory\cite{Altshuler} to find the change in the resonant frequency:  
\begin{equation}
\frac{\Delta f}{f} \approx \frac {\epsilon _{0} (\epsilon _{r2} - \epsilon _{r1} )} {4W} 
\int _{V _{S}} \textbf{E} _{1} \cdot \textbf{E} _{2} dV ,
\label{delta_f_equation}
\end{equation}
where $\bf E$$_{1}$ and $\bf E$$_{2}$, and $\epsilon _{r1}$ and $\epsilon _{r2}$ are the unperturbed and perturbed electric fields, and relative permittivities of two samples, respectively, $W$ is the energy stored in the resonator, and the integral is over the volume of the sample. We compute an approximate $W$ using the fact that the loaded quality factor\cite{Steinhauer2} of the resonator is $Q _{L} = \omega _{0} W/P _{loss}$, where $\omega_{0}$ is the resonant frequency, and $P _{loss}$ is the power loss in the resonator. Using four bulk samples with known relative dielectric permittivities between 2.1 and 305, and fixing $r _{0} = (0.6\ \mu$m$) / \alpha$, we used $\alpha$ as a fitting parameter to obtain agreement between the model results from Eq.\ (\ref{delta_f_equation}) and our data at 7.2 GHz (see the graph in Fig.\ \ref{schematic}); we found agreement to within 10 \% for several different probe tips, with $1.0 < \alpha < 1.7$.

To extend this calibration model to thin films, we extend the finite element calculation to include a thin film on top of the dielectric sample substrate.  Once the $\alpha$ parameter of a probe is determined using the bulk calibration described above, we use the thin film model combined with Eq.\ (\ref{delta_f_equation}), integrating over the volume of the thin film, to obtain a functional relationship between $\Delta f$ and the dielectric permittivity of the thin film. Using the thin film model, we found that for high-permittivity ($\epsilon_{r} \gtrsim 50$) thin films, the microwave microscope is primarily sensitive to the in-plane component of the permittivity tensor.

\begin{figure}
\begin{center}
\leavevmode
\epsfxsize=18cm
\epsffile{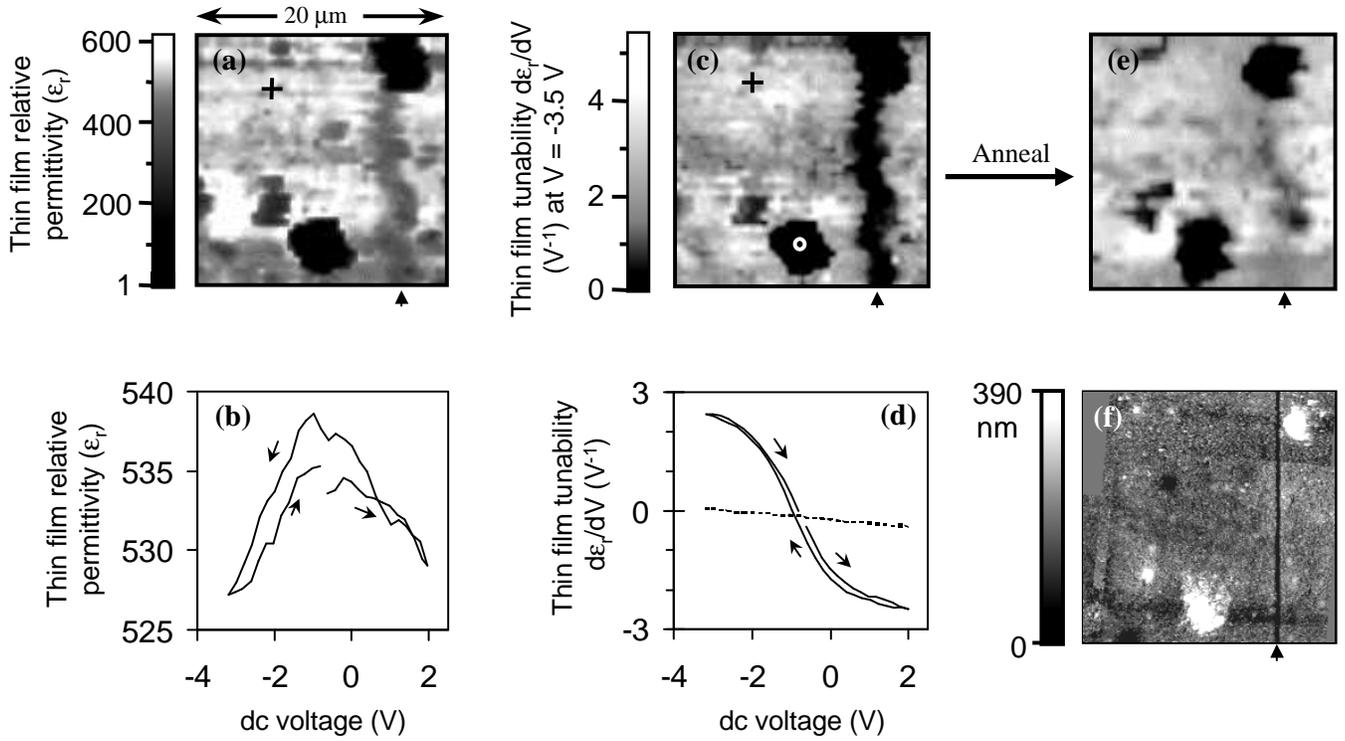}
\end{center}
\caption{Images and hysteresis loops at 7.2 GHz of the BST/LSCO/LAO thin-film sample (see inset to Fig.\ \ref{schematic}). All images show the same 20 $\times$ 20 $\mu$m$^{2}$ area of the sample.  (a) Permittivity image. (b) Hysteresis loop taken at the location marked "+" in (a). (c) Dielectric tunability image. (d) Dielectric tunability hysteresis loops.  The solid line corresponds to the "+" in (c), while the dashed line corresponds to the "$\circ$" in (c).  (e) Dielectric tunability image taken after the sample was annealed at 650 $^{\circ}$C in air for 20 minutes; note the change from (c).  (f) Atomic force microscope topographic image.  The milled vertical line marked by the arrow in (f) appears rough in (a) and (c) because of drift in the microscope during scanning.}
\label{images}
\end{figure}

Figure \ref{images}(a) shows a quantitative permittivity $\epsilon_{r}$ image of a sample (see inset to Fig.\ \ref{schematic}) consisting of a 370 nm Ba$_{0.6}$Sr$_{0.4}$TiO$_{3}$ (BST) thin film on a 70 nm La$_{0.95}$Sr$_{0.05}$CoO$_{3}$ (LSCO) counterelectrode.  The substrate is LaAlO$_{3}$ (LAO).  The films were made by pulsed laser deposition at 700$^{\circ}$ C, in 200 mTorr of O$_{2}$.  The film is paraelectric at room temperature ($T_{c} \sim 250$ K). We measured the film thickness using scanning ion microscope images of cross-sections milled by focused ion beam (FIB).  To identify the scanned area on the sample, a 0.1 $\mu$m wide line was FIB milled through the BST layer.  This fiducial mark allowed further analysis of the area of interest with an atomic force microscope (AFM) [Fig.\ \ref{images}(f)].  In the microwave permittivity image [Fig. \ref{images}(a)], the milled line (marked by an arrow) is visible as a low-permittivity vertical band.  Two low-permittivity defects on the film are also visible. The average value of the permittivity of the film far from any defects is $\epsilon_{r} = 510$, in reasonable agreement with typical measurements of $\epsilon_{r}$ in similar films.\cite{WChang,QXJia}

In order to measure the local dielectric tunability of thin films, we apply a dc electric field to the sample by voltage biasing ($V_{bias}$) the probe tip (see Fig.\ \ref{schematic}).  A grounded metallic counterelectrode layer immediately beneath the dielectric thin film acts as a ground plane.  To prevent the counterelectrode from dominating the microwave measurement, thus minimizing its effect on the microwave fields (we ignored the counterelectrode in our static field model), the sheet resistance of the counterelectrode should be as high as possible.  In our case, we use low carrier density LSCO with a thickness of 70 nm, giving a sheet resistance $\sim 400\ \Omega \slash \Box$.  We have confirmed by experiment and model calculation\cite{Steinhauer1} that the contribution of the counterelectrode to the frequency shift is small ($\Delta f < 30$ kHz) relative to the contribution from a dielectric thin film with thickness $>$ 100 nm ($\Delta f > 200$ kHz).

We first examined the tunability of the dielectric properties at a fixed point on the sample (marked by a "+" in Fig.\ \ref{images}(a); the resulting hysteresis loop is shown in Fig.\ \ref{images}(b)).  As expected, the permittivity goes down when a voltage is applied.  The inverted "V" curve is centered at a bias of -0.6 V rather than 0 V probably because the asymmetric electrodes (a thin film on the bottom, and a sharp tip on top) induce unequal charges on the two surfaces.\cite{Miranda}  The tunability of $\sim 2$\% is small compared to measurements on similar films (20-30\%) using standard techniques.\cite{Hoerman,Miranda}  The cause of this low tunability is unclear, but may be due to the microscope measuring an off-diagonal component of the nonlinear permittivity ($\epsilon_{r}$ measured in the horizontal direction, while the applied field is in the vertical direction). Other explanations may be the existence of a low-permittivity depletion layer above the counterelectrode, or the microwave measurement averaging over a larger area than the area in which the dc electric field is concentrated.

To image the tunability of the sample, we modulate the bias voltage applied to the probe tip at a frequency $\omega_{bias} =$ 1 kHz and amplitude $\pm1$ V (see inset to Fig.\ \ref{schematic}), and monitor the component of the microscope frequency shift signal at $\omega_{bias}$ with a lock-in amplifier.  This $\omega_{bias}$ component is proportional to $d\epsilon_{r}/dV$, where $V$ is the applied voltage.  An image of $d\epsilon_{r}/dV$ at $V = -3.5$ V is shown in Fig.\ \ref{images}(c).  We note that the tunability diminishes near the FIB milled line and two large defects.  Figure \ref{images}(d) shows hysteresis loops at a high-permittivity location and at a low-permittivity defect, marked by the "+" and "$\circ$" symbols in (c), corresponding to the solid and dashed lines in (d), respectively.  As shown in (c) and (d), the tunability at the defect is small relative to the rest of the film.  The small amount of hysteresis observed is probably due to the broadened ferroelectric transition in thin films, which may be due to variations in stoichiometry and the transition temperature.  If desired, the data in Fig.\ \ref{images}(c) could be converted into an approximate fractional tunability ($\Delta \epsilon_{r}/\epsilon_{r}$) image.

After annealing the sample at 650 $^{\circ}$C in air for 20 minutes, we scanned the sample again, as shown in the $d\epsilon_{r}/dV$ image in Fig.\ \ref{images}(e).  The milled line, which is prominent in Fig.\ \ref{images}(c), is nearly invisible in (e), indicating that the dielectric tunability near the milled line has been almost completely restored.  

The AFM image in Fig.\ \ref{images}(f) shows many of the features seen in the microwave microscope images (a) and (c).  The gallium FIB milled vertical line is visible, as well as the two large low-permittivity defects, which are shown to be particles on the surface in (f).  These defects are probably non-vapor phase particles which accrued during pulsed laser deposition.  They appear slightly larger in the microwave microscope images, indicating that the spatial resolution of the microscope in this case is about 1 $\mu$m.  The dark horizontal bands next to the large defects are artifacts of the AFM imaging technique.  We also note that the AFM image, which was acquired after the microwave images, does not show any evidence of scratching by the microwave microscope probe tip.

The microwave technique we use is sensitive to both film thickness and permittivity.  As a result, the permittivity of the two large defects in Fig.\ \ref{images}(a) is underestimated due to the change in film thickness at these locations, which we have confirmed with our model.  However, by examining the AFM image [Fig.\ \ref{images}(f)], topographic features can be readily distinguished from permittivity features. Far from large topographic features in the images in Fig.\ \ref{images}, the AFM data indicates an rms surface roughness of about 2 nm (when the data is averaged over the microwave microscope spatial resolution of 1 $\mu$m), which would result in an error of $\Delta \epsilon_{r} \sim 1.5$ in the permittivity data. This is much less than the observed variation $\Delta \epsilon_{r} \sim 30$ in the same area, indicating that this observed variation in $\epsilon_{r}$ is mainly due to permittivity contrast.

We can calculate the sensitivity of the microwave microscope by observing the noise in the dielectric permittivity and tunability data.  For a 370 nm thick film on a 500 $\mu$m thick LAO substrate, with an averaging time of 40 ms, we find that the relative dielectric permittivity sensitivity is $\Delta\epsilon_{r} = 2$ at $\epsilon_{r} = 500$, and the tunability sensitivity is $\Delta (d\epsilon_{r}/dV) = 0.03$ V$^{-1}$.

In conclusion, we have demonstrated the use of a near-field scanning microwave microscope to quantitatively image the local permittivity and tunability of dielectric thin films with a spatial resolution of 1 $\mu$m. This method is nondestructive and has broadband (0.1 to 50 GHz) capability.

This work has been supported by NSF-MRSEC grant No.\ DMR-9632521, NSF grants No.\ ECS-9632811 and DMR-9624021, and by the Maryland Center for Superconductivity Research.

\newpage

\newpage

\end{document}